\documentclass[10pt,twocolumn,conference]{IEEEtran}
\IEEEoverridecommandlockouts

\def\@IEEEpubidpullup{1.3\baselineskip}
\makeatother

\usepackage{amsfonts}
\usepackage{times}
\usepackage[english]{babel}
\usepackage{times}
\usepackage{bbold}
\usepackage{bbm}
\usepackage{url}
\usepackage{verbatim}

\usepackage{dsfont}

\usepackage{indentfirst}
\usepackage{cite}
\usepackage{subfigure}
\usepackage{amsmath}
\usepackage{multicol}
\usepackage{amsfonts}
\usepackage{times}
\usepackage{bm}
\usepackage{mathrsfs}
\usepackage[dvips]{graphicx}
\usepackage[usenames]{color}

\usepackage{amssymb}

\newcommand{\figwww}{0.83\columnwidth}

\newcommand{\bi}{\begin{itemize}}
\newcommand{\ei}{\end{itemize}}
\newcommand{\ben}{\begin{enumerate}}
\newcommand{\een}{\end{enumerate}}
\newcommand{\bc}{\begin{cases}}
\newcommand{\ec}{\end{cases}}
\newcommand{\bd}{\begin{description}}
\newcommand{\ed}{\end{description}}

\newcommand{\be}{\begin{equation}}
\newcommand{\ee}{\end{equation}}
\newcommand{\bea}{\begin{eqnarray}}
\newcommand{\eea}{\end{eqnarray}}

\graphicspath{{./figures/}{./images/}}

\begin{document}

%\sloppy

\title{Supporting Multi-hop Device-to-Device Networks Through WiFi Direct Multi-group Networking}

\author{\IEEEauthorblockN{Colin Funai, Cristiano Tapparello, Wendi Heinzelman}\\
Department of Electrical and Computer Engineering, University of Rochester,\\
Rochester, NY, USA (firstname.lastname@rochester.edu)
\thanks{This work was supported in part by Harris Corporation, RF Communications Division and in part by CEIS, an Empire State Development designated Center for Advanced Technology.}}
\maketitle
%\thispagestyle{empty}

% ABSTRACT
\begin{abstract}
%\section{abstract}
\label{sec:abstract}
With the increasing availability of mobile devices that natively support Device-to-Device (D2D) communication protocols, we are presented with a unique opportunity to realize large scale ad hoc wireless networks. Recently, a novel D2D protocol named WiFi Direct has been proposed and standardized by the WiFi Alliance with the objective of facilitating the interconnection of nearby devices. However, WiFi Direct has been designed following a client-server hierarchical architecture, where a single device manages all the communications within a group of devices. 
In this paper, we propose and analyze different solutions for supporting the communications between multiple WiFi Direct groups using Android OS devices. By describing the WiFi Direct standard and the limitations of the current implementation of the Android WiFi Direct framework, we present possible solutions to interconnect different groups to create multi-hop ad hoc networks. 
Experimental results show that our proposed approaches are feasible with different overhead in terms of energy consumption and delay at the gateway node. Additionally, our experimental results demonstrate the superiority of techniques that exploit the device ability to maintain simultaneous physical connections to multiple groups.
 
\end{abstract}

\section{Introduction}
\label{sec:introduction}

The growing popularity of mobile devices such as tablets and smartphones, combined with their native support of device-to-device (D2D) communication protocols, has brought forth research and development of new ad hoc communication techniques. For instance, starting from Android 4.0, Google has introduced a WiFi peer to peer framework~\cite{AndroidWiFip2p} to its mobile operating system, while Apple has introduced a framework for enabling D2D mesh networking among iOS devices~\cite{iOS}. The idea of extending the cellular infrastructure through D2D communication has also been proposed, and a standard for allowing direct communication between nearby mobile devices will be added to LTE in 3GPP Release 12~\cite{3GPP}. It is easy to imagine a future where cellular service providers capitalize on D2D communications to extend the cellular coverage.
%However, whether this will actually benefit the user's normal experience remains to be explored at this time, as well as whether further evolutions into Mobile Ad Hoc Networks (MANETs) are feasible without impacting the final users performance. 

The diffusion of these new paradigms for D2D communication has in turn encouraged researchers to investigate their impact on the users' experience. Several studies showed that using a D2D communication technology has the potential to improve both the spectral efficiency and the device battery life while, at the same time, providing better resource utilization for infrastructure-communications protocols like WiFi~\cite{Asadi2014}. 

Over the years there have been many attempts to enable direct communications between IEEE 802.11 radio devices, such as IEEE 802.11 DCF, 802.11s, and 802.11z. While these protocols each have their application areas, their diffusion is quite limited. Recently, a new IEEE 802.11 based protocol named WiFi Direct~\cite{WiFip2p} has been released by the WiFi Alliance to address the shortcomings of IEEE 802.11 DCF. WiFi Direct is now natively included in most of the modern mobile devices and, due to its diffusion and promising performance, it is receiving considerable attention from the research community~\cite{Camps2011,Mur2013,Conti2013}.
WiFi Direct aims to enhance WiFi based D2D communications~\cite{Mur2013} and has been designed with energy saving mechanisms leading to higher energy efficiency. With WiFi Direct, devices are organized in groups, where one member of the group is the Group Owner (GO) and all the other devices are Group Members (GM). Being built on top of WiFi, groups are able to also support legacy clients, devices that do not support WiFi Direct but support WiFi. 

Multi-hop wireless networks have been largely developed to meet the needs of a variety of applications where infrastructure-based wireless networks are difficult to deploy and maintain. Most applications require the participating nodes to be able to route data to help extend network connectivity. These protocols have mainly been used for tactical military communications, first responder applications and sensor network operations.

Given the wide availability of WiFi Direct devices, multi-group WiFi Direct communications can be used to create multi-hop ad hoc networks, greatly benefiting many scenarios where the communication infrastructure is either overloaded, damaged or not present.  
However, the WiFi Direct standard~\cite{WiFip2p} defines only intragroup communications, with the GO being at the center of all the communications, but does not preclude a WiFi Direct device to simultaneously operate as a member of more than one group. 

In this paper, we explore different methods for allowing the communication among devices that belong to different WiFi Direct groups. Starting with stock Android and its implementation of WiFi Direct, we shed some light on the limitations and design considerations of realizing multi group communication on mobile devices running Android $4.4.2$. To overcome these limitations, we first propose and analyze a TCP-based time sharing mechanism where the device that connects multiple groups, referred to as the gateway node, is required to iteratively switch between different WiFi Direct groups in order to relay data from one group to the other. We then exploit a particular configuration that allows for a device to be simultaneously connected to two different groups, to implement and analyze the performance of a UDP-based broadcast technique as well as a UDP/TCP hybrid solution. In addition, we consider how to implement intergroup communication when not limited to a stock version of Android and present a few changes that can be made to the current WiFi Direct implementation in order to facilitate intergroup communication. Finally, we discuss the tradeoffs in terms of both energy and time when implementing these different approaches. Our approaches can be used as building blocks for realizing a WiFi Direct based Mobile Ad Hoc Network by interconnecting Android devices. 

%It is important to note the work done by Duan et al.~\cite{Duan2014}, which  have proposed a method of facilitaing multi-group communication and a more detailed comparison of our approaches will be addressed in ~\ref{sec:AndroidMultiGroup}.
A similar idea has recently been proposed in~\cite{Casetti2014}. The authors in~\cite{Casetti2014} propose a multi-group data dissemination protocol in which Android devices belonging to different groups collaborate with each other by broadcasting UDP packets, as in our UDP-based solution. However, the protocol presented in~\cite{Casetti2014} is based on a particular logical topology that requires, in addition to the gateway node, the presence in every group of an additional relay node. We will examine the benefits and issues with this approach in Section~\ref{sec:proposed_solutions}.

%This in turn would require a TCP-over-UDP~\cite{dunigan2002} like implementation in order to achieve reliability and can negatively impact performance as detailed in~\cite{dunigan2002,Bruno2008,Friedman2009}.

The rest of the paper is organized as follows. Section~\ref{sec:WiFiDirect} provides some background on the WiFi Direct standard. 
Section~\ref{sec:AndroidMultiGroup} briefly describes the current Android implementation of WiFi Direct, its limitations, and our solutions to realize multi-group networking with Android devices. In Section~\ref{sec:results} we discuss our experimental results, highlighting the energy consumption and evaluating the delay of the different methods considered. Section~\ref{sec:conclusions} concludes the paper.

% Talk about the multi groups extension of WiFi Direct
%Cite other approaches

%Camps-Mur et al. have proposed to use the Notice of Absence (NoA) to reduce the power consumption at the GO. Notice of Absence is a power saving mechanism that allows for a node to power down its radio in an attempt to save power~\cite{Mur2013}.
%NoA is a mechanism where a node requests to be gone from the group for a set period of time in order to power down the radio and save energy. 
%After the set time, the node must return to the group and alert the GO that they are still alive.
%Camps et al. propose that rather than powering the radio down, the time that the requesting node is allowed to be away can be used to service another group.
%

\section{WiFi Direct}
\label{sec:WiFiDirect}

\subsection{Single-group Communications}
%background on WiFi direct
WiFi Direct~\cite{WiFip2p} is a standard released by the WiFi alliance that enables D2D communication between nearby devices, without requiring a wireless Access Point (AP). WiFi Direct utilizes IEEE 802.11 a/b/g/n infrastructure mode, and can transmit either at 2.4~GHz or 5~GHz.  

During D2D communication, devices form a group were one of them is the Group Owner (GO) and all the others are considered Group Members (GM). It is important to note that these roles are not predefined but are negotiated during the construction of the group and remain fixed for the entire duration of the group. Additionally, WiFi Direct groups can also include standard IEEE 802.11 nodes that do not support WiFi Direct and are referred to as Legacy Clients (LC). %An example scenario of a WiFi Direct group is presented in Figure~\ref{fig:scenarioWiFiDirect}.

%\begin{figure}
 %   \centering \includegraphics[width=0.7\columnwidth]{./figs/scenarioWiFiDirect2.eps}
 %   \caption{Example scenario of a WiFi Direct communication group, with corresponding functional roles. Group Owner (GO), Group Member (GM) and Legacy Client (LC).}
%    \label{fig:scenarioWiFiDirect}
%\end{figure}

The nodes that support WiFi Direct go through a group formation process in order to determine the roles of the GO and the GMs. There are three group formation cases: standard, persistent and autonomous~\cite{WiFip2p},~\cite{Mur2013}. 
During the standard group formation, the nodes listen on channels $1$, $6$, and $11$ in the 2.4~GHz band and, after finding another device, they negotiate as to which will act as the GO. This is done in a handshake process, where the devices exchange an \textit{intent value}, and the device with the highest value becomes the GO. After the roles have been established, the devices go through a WiFi Protected Setup (WPS) Provision phase and, after completion, the GO assigns the IP address using the Dynamic Host Configuration Protocol (DHCP).
The persistent group formation process allows for a faster reconstruction of previous groups. During the persistent group formation, the GO negotiation phase is replaced by an invitation exchange, and the WPS Provisioning process is significantly reduced by reusing the stored network credentials. 
In autonomous group formation, a node assigns itself the role of GO and creates its own group.

According to the standard~\cite{WiFip2p}, the GO represents an AP-like entity that provides basic service set (BSS) functionality and services for the associated clients. Acting as a soft AP, the GO advertises and allows nodes to join the group. The advertisement and group maintenance are performed through beacon packets, just like a typical IEEE 802.11 AP, and the GO is responsible for giving control of the channel to nodes in its network as well as routing data through clients in its group\footnote{Routing data between clients in a group is allowed but not defined by the standard.}. As a result, the group topology is a $1:N$ hierarchical structure, where multiple clients (i.e., GMs and LCs) are connected to one GO.% (see Figure~\ref{fig:scenarioWiFiDirect}). 

WiFi Direct devices can operate concurrently with an infrastructure wireless network, through multiple physical or virtual MAC entities. Moreover, the specification~\cite{WiFip2p} does not preclude a WiFi Direct device from simultaneously operating as a member of more than one group. However, both the multiple MAC functionalities and the simultaneous operations in multiple groups are out of scope of the standard.

\subsection{Multi-group Communications}
\label{WiFiDirectMulti}

Our focus is to investigate the feasibility and relative performance of different techniques for allowing communication between different WiFi Direct groups. In this regard, in order to act as a gateway between two (or more) WiFi Direct groups, a device can use the MAC virtualization functionality described earlier. Thus, the physical radio interface can be shared by multiple separate MAC entities that independently use the hardware. Following the same principle, a device can act as a gateway between a local D2D network and the Internet, through the simultaneous connection to an infrastructure AP. We note that, when connected to a standard WiFi AP, the device is in fact acting as a LC since it is not leveraging the WiFi Direct protocol. 

\begin{figure}
    \centering \includegraphics[width=0.65\columnwidth]{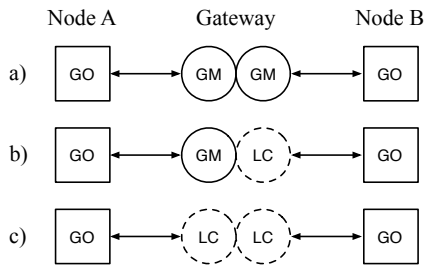}
    \caption{Multi-group communication scenarios where the gateway node acts as a client in two groups.}
    \label{fig:scenarioGM}
\end{figure}

\begin{figure}
    \centering \includegraphics[width=0.65\columnwidth]{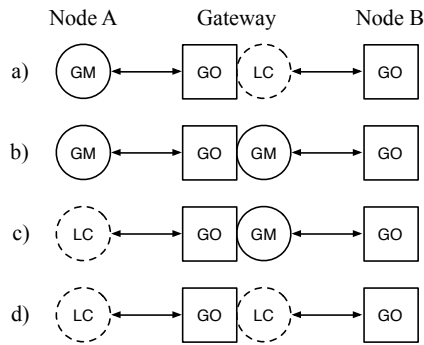}
    \caption{Multi-group communication scenarios where the gateway node acts as the GO in one group and as a client in the other.}
    \label{fig:scenarioGO}
\end{figure}

Given the above, we envision two possible scenarios in which a device can act as a gateway between two separate groups: the first where the gateway node acts as a client in both groups (see Figure~\ref{fig:scenarioGM}), and the second scenario in which the gateway is the GO of one group and a client in the other (see Figure~\ref{fig:scenarioGO}). Extensions of these scenarios to more than two groups or the case in which the GO hosts more than one group are also possible. However, the case in which the GO hosts more than one group only allows for an increase in the number of clients that the GO can simultaneously serve. 

\section{Multi-group Networking on Android Devices}
\label{sec:AndroidMultiGroup}

As described in the previous section, a traditional WiFi Direct network topology is represented by a hierarchical structure, where the GO is at the center of all the communications, but multi-group communications are allowed by the standard specification. In principle, it is therefore possible to realize a multi-group wireless network where some of the devices are clients, or simultaneously a GO and a client, of more than one group. However, the MAC virtualization and the simultaneous operations in multiple groups are not required by the standard, and thus their availability depends on the actual implementation. 

In what follows, we first provide a high level description of the Android implementation of WiFi Direct, we then present the limitations of using stock Android, and we finally describe our proposed methods for realizing intergroup communications with stock and non-stock Android devices. 

\vspace{-1mm}
\subsection{WiFi Direct}
\vspace{-1mm}

Android has included support for WiFi Direct since version $4.0$ (API level 14), within the Android's Wi-Fi P2P framework~\cite{AndroidWiFip2p}. This framework complies with the WiFi Alliance's WiFi Direct certification program. Using the Android APIs, a developer can discover and connect to other devices that support WiFi Direct and then communicate over a D2D connection. 
The {\it WifiP2pManager} class provides all the methods that allow the interaction with the WiFi hardware, like discover, connect and disconnect to peers. Due to the interaction with the hardware, all these methods are asynchronous and the framework uses listeners to notify the application of the status of a call.

In order to use the {\it WifiP2pManager} functionalities, an Android application needs to have access to the hardware and run on a device that supports the WiFi Direct protocol. If both these conditions are satisfied, the {\it WifiP2pManager} undergoes an initialization process, where all the WiFi Direct related services are started. This allows the device to react to WiFi Direct events, like client discovery and connection. For establishing a D2D connection, a device needs to discover nearby peers that support WiFi Direct and are available for connection. After a device is discovered, the actual connection can be made. If one of the two devices is already a GO of an existing group, the other joins the group as a GM, while if none of them is a GO or a GM, according to the WiFi Direct protocol described earlier, the devices negotiate their role within the group. Moreover, if a GO sends a connection request to the GM of another group, the connection is refused, and if a device sends a request to a GM, the GM forwards the request to its GO. We note that a device can be a GO of a group without any connected clients. 

After the connection, the GO acts as a Dynamic Host Configuration Protocol (DHCP) server and assigns an IP address to the clients. The DHCP scope is fixed by the framework and cannot be modified by the developers. As a result, GOs of different groups always have the same IP address ($192.168.49.1$), while the connected clients receive an IP address  at random from the same range ($192.168.49.2$-$254$).

The GO advertises its group through a unique Service Set ID (SSID), that can be used by devices that do not support the framework (i.e., LC) to join the group. In Android, standard WiFi operations, like scanning for available networks and the connection to infrastructure APs, are handled by the WiFi framework. Developers can use the {\it WifiManager} class to perform WiFi specific functionalities. 

For a detailed description of the WiFi and WiFi Direct frameworks and the relative APIs, we refer interested readers to the Android API guides~\cite{AndroidAPI}.

\vspace{-1mm}
\subsection{Limitations of Stock Android}
\vspace{-1mm}
\label{sec:limitations}
Even though Android is an open-source operating system, it provides a certain set of limitations in the way in which the developer can interact with the different services and hardware. We acknowledge that these limitations can be removed by reprogramming the operating system (e.g., rooting the device). However, this operation is non trivial for an average user and its legality is still controversial in several countries. Thus, we first focus on devices that run stock Android so that we can provide a general methodology for allowing WiFi Direct multi-group networking on a broader set of devices. 

Using stock Android, intergroup communications need to be handled at the application layer, and all the transport and network layer functionalities, like setting the IP address and managing the routing table, cannot be performed natively. Moreover, the developer is not allowed to create either custom virtual network interfaces nor multiple virtual MAC entities. As a result, the methods described in Section~\ref{WiFiDirectMulti}, where a device simultaneously operates in multiple WiFi Direct groups either as a GO or a GM, cannot be implemented directly.% on Android. 

Nevertheless, our experiments show that the WiFi Direct functionalities are able to concurrently operate with an infrastructure wireless network\footnote{We note that this functionality is not described in the Android APIs~\cite{AndroidAPI}.}, through the simultaneous utilization of the {\it WifiP2pManager} and {\it WifiManager}. In this case, we infer that the OS is in fact virtualizing the network interface (or the MAC). Following the same rationale, we exploited the GO group advertisement process and connected a device participating in a WiFi Direct group to a second group as a LC, using the {\it WiFiManager}. While this operation was allowed and the connection between the devices was correctly established, we were not able to create a unicast communication to and from the gateway node. In particular, for the scenario b) of Figure~\ref{fig:scenarioGM}, the gateway was able to receive data from both Node A and Node B, but was not able to communicate with either one of them; for the scenarios a) and d) of Figure~\ref{fig:scenarioGO}, instead, the gateway was able to communicate with Node A, while the communication with Node B was not allowed. According to our experiments, this is due to the fact that the DHCP protocol assigns the same IP address to multiple GOs, thus creating routing problems. A flooding like UDP-based communication protocol can overcome this limitation, as described in Section~\ref{sec:proposed_solutions}.

\vspace{-1mm}
\subsection{Proposed Solutions}
\vspace{-1mm}
\label{sec:proposed_solutions}
{\it Time Sharing.} 
Given the limitations of the current Android implementation of WiFi Direct and the restrictions on implementing routing functionalities at the application layer, we propose a time sharing mechanism in which the gateway node switches between two (or more) groups. In this way, all the scenarios presented in Figures~\ref{fig:scenarioGM} and~\ref{fig:scenarioGO} can be successfully implemented. We note that there is no built in switching functionality, rather switching is comprised of a disconnection from the current group, a request to scan for active nodes, and a request to connect to a new group. In what follows, we describe the main differences between the different scenarios. 

We start by analyzing the scenarios in Figure~\ref{fig:scenarioGM}. In these cases, the gateway node acts as a client in both groups and iteratively connects to and disconnects from the two groups. Scenarios a) and b) are limited to devices that support WiFi Direct: while in case a) the gateway can fully exploit the WiFi Direct protocol, the latter can potentially provide some gains in terms of switching time since it is using both the {\it WiFiManager} and the {\it WiFiP2pManager}. A special case is represented by case c), where the gateway is a LC in both groups. This method relies solely on the {\it WiFiManager}, thus allowing devices that do not support WiFi Direct to take on the onerous role of routing information between different groups. By acting as a LC, the gateway cannot capitalize on the WiFi Direct power saving mechanisms, but it follows the same process required for switching connections between traditional WiFi APs. While methods described in a) and b) have the potential for a faster and seamless bridging of groups, the method in c) allows for any device to act as a gateway node, allowing for a more inclusive network. 

The scenarios in Figure~\ref{fig:scenarioGO}, instead, consider the situation in which the gateway node is a client in one group and acts as a GO of the other. In these cases, the switching process potentially requires more time then the methods described in Figure~\ref{fig:scenarioGM} because all of the operations required to create one of the groups need to be performed every time the gateway node ceases to be the GO. The role and responsibilities of being a GO are negotiated during the group formation and cannot be transferred~\cite{WiFip2p}. Scenarios b) and c) in Figure~\ref{fig:scenarioGO} fully exploit the WiFi Direct protocol, while scenarios a) and d) use both WiFi Direct and the standard WiFi functionalities.

{\it Simultaneous Connections.} 
As described in Section~\ref{sec:limitations}, by combining standard WiFi and WiFi Direct functionalities into the gateway node, it is possible for the gateway node to simultaneously maintain a physical/MAC layer connection to two groups. However, this imposes some restrictions on the actual application data exchange. We thus explored the different scenarios presented in Figures~\ref{fig:scenarioGM} and~\ref{fig:scenarioGO} that utilize both WiFi Direct and WiFi, with the objective of finding a suitable configuration and communication protocol that allow for intergroup data exchange.   

To this end, we ran several experiments by implementing the different scenarios from Figures~\ref{fig:scenarioGM} and~\ref{fig:scenarioGO} with the different network sockets (e.g, stream, datagram and multicast sockets) provided by Android. We found that when combining a {\it LC/GM} (or, equivalently, {\it GM/LC}) gateway node with a multicast socket, a specific implementation of the UDP datagram socket, the gateway node is able to forward data between the two groups. This is because the multicast socket allows the node to specify the particular interface to be used by the socket for receiving and transmitting data packets. We note that this functionality is not available for a traditional datagram or stream socket, and Android follows a weak end system model (i.e., routing decision are based only on the destination IP address and type of service)~\cite{RFC1122}. Moreover, it is important to note that the multicast socket encapsulates a one-to-many unicast communication and, as a result of this, cannot fully utilize the total available WiFi and WiFi Direct bandwidth. 

%although we use a multicast socket, this is not true multicast, rather this is many-to-one unicast and {\it GO} acts as the coordinator. This means that the reception from the WiFi Direct link is very slow since the {\it GO} has to constantly look up which nodes are active in this multicast group. 

From our experiments, the same gateway configuration allows the gateway node to receive and send data over the {\it LC} link (i.e., the standard WiFi) while simultaneously connected to both groups also with a unicast socket, while no data can be routed with a unicast socket over the WiFi Direct link. This is due to the fact that Android prioritizes the WiFi link over the WiFi Direct link. We thus propose a simple protocol, referred to as {\it Hybrid}, that exploits this functionality and uses the multicast socket as a control channel that, if necessary, triggers a gateway node configuration change. According to this protocol, the group that has data to send to the other group, uses the control channel to notify the gateway node. After the reception of the control message, the gateway node checks if it is connected to this group as a {\it LC} or a {\it GM}. In the first case, the gateway is allowed to receive data, thus it notifies the source and it starts receiving data. After receiving the data, the gateway disconnects from the first group and forwards the data to the second group using a TCP connection. In the second case, instead, the gateway node is not allowed to receive data from the WiFi Direct link. Thus, it notifies the source, disconnects from both groups and connects back with the right configuration (i.e., it switches from {\it GM/LC} to {\it LC/GM}). At this point the communication continues as in the previous case. We note that this change in configuration can be avoided if a second gateway node is present (i.e., one gateway node is a {\it GM/LC} while the other gateway node is an {\it LC/GM}).

A data dissemination protocol that follows a similar approach has recently been proposed in~\cite{Casetti2014}. The protocol presented in~\cite{Casetti2014} operates under the assumption that an additional relay node is present in each group. This protocol uses the configurations a) and d) of Figure~\ref{fig:scenarioGO}, and exploits the additional relay node to forward information from the GO to the gateway node. By doing so, the method presented in~\cite{Casetti2014} mitigates the fact that the gateway node has the same IP address as its GO. However, this does not provide seamless communication between two groups, since a TCP connection defaults to the WiFi interface (i.e., only the LC link can use TCP sockets). As a result, the gateway node transmits data to the other nodes in its group by broadcasting UDP packets. While this protocol does not require any configuration change, it requires the presence of an additional relay node in every group (which cannot always be guaranteed), and adds an additional communication hop.% when routing data between groups.  

In order to fully explore the WiFi Direct protocol for multi-group networking, we  downloaded the source code of Android $4.4.2$ and modified the existing implementation of WiFi Direct to assign a unique IP address to the GOs and change the DHCP range accordingly (the IP address and DCHP range are statically defined inside the {\it WifiP2pService}). This simple modification allows the simultaneous operation of a gateway node that uses both the WiFi and WiFi Direct interfaces. Further modifications are required for creating multiple WiFi Direct interfaces. This requires changing the {\it WifiP2pService} to instantiate different {\it NetworkInterfaces} as well as to change the {\it SystemServer} and the {\it Context} to instantiate a new WiFi Direct Service and a new identifier for this service, respectively.

%This can be done by modifying the constructor for the {\it WifiP2pService} which specificies the {\it NetworkInterface} to be {\it p2p0} by default. An additional interface can be created by changing this to check to see if {\it p2p0} exists and takeing appropriate actions if it does. This technique also requires changes to be done in {\it SystemServer.java} to instantiate a new WiFi Direct Service, as well as changing {\it Context.java} to add a new identifier for this service. 

\section{Performance Evaluation}
\label{sec:results}

In this section we evaluate the impact at the gateway node of allowing multi-group WiFi Direct communications. In this regard, we measured both the time and the energy required to forward data between two groups. Our proposed schemes were implemented using three second generation Asus Nexus 7s, which were released in 2013. The 2013 Nexus 7 has 16 GB of storage, 2 GB of memory, and a 1.5GHz quad-core Snapdragon S4 Pro 8064 CPU~\cite{Nexus7}. Each of our devices are running Google's Android version $4.4.2$.% produced by Qualcomm. 

\vspace{-1mm}
\subsection{Test Environment}
\vspace{-1mm}
Similar to the work in~\cite{Friedman2009}, we measured the current from the battery for each of our experiments using an Arduino Uno~\cite{arduino}. To this end, we added a $0.005 \Omega \pm 1\%$ shunt resistor in series with the battery and measured the voltage across the resistor to obtain the current. However, the voltage drop across this resistor was too small to be read by the Arduino Uno~\cite{arduino}. We therefore configured an op-amp to act as a non-inverting amplifier with a gain of $977$. This allowed the Arduino's analog input to read the voltage across the resistor throughout the tests. Additionally, due to the non-linear properties of lithium-ion batteries~\cite{Rak2003}~\cite{coleman2007}, like the one used in the Nexus 7, we restricted our experimental measurements to battery level above $70$\%. \cite{coleman2007} has shown that for battery levels between $60$\% and $100$\%, the internal impedance of lithium-ion batteries scales almost linearly with regard to the state of charge. 

%\begin{figure}
 %   \centering \includegraphics[width=0.875\columnwidth]{./figs/powerVStime_GM_GOGM_GO_labels.eps}
 %   \caption{Example of the power readings for a single experiment. A $10$~s delay has been added between each phase of the experiment.}
 %   \label{fig:sample}
%\end{figure}

All of the measurements were taken in an indoor workspace, and the nodes were stationary during these experiments. We acknowledge that in real life, nodes are generally mobile, and may enter or leave a network. However, to accurately determine the impact of the switching process on the gateway node and to limit the variability across different experiments, we kept the nodes stationary. Moreover, all the results of this section have been obtained with the screen turned off.

For all the scenarios presented in Figures~\ref{fig:scenarioGM} and~\ref{fig:scenarioGO}, we consider a situation in which the gateway must relay $10$~MB of data between Node A and Node B. Thus, the gateway first receives all the data from one node, and then sends the data to the other node following one of the techniques described in Section~\ref{sec:proposed_solutions}. For the {\it Time Sharing} experiments, we measured the energy and time required for the switching process, from the disconnection from the first group to the actual availability of the second link. We assumed that the gateway is able to communicate with the second node as soon as the operating system updates the ARP table with its address. For the {\it Simultaneous Connections} solutions, instead, we measured the total time and energy consumption required to relay the data between node A and Node B, since there is no actual switching between groups. Moreover, all the groups are considered to be persistent to allow for a faster switching between the two groups, that we assume to be known by the gateway. This accounts for the situation in which the gateway node switches between the groups multiple times\footnote{We acknowledge that, when creating the groups for the first time, the energy and time required are on average higher.}. All the results presented in this section are obtained over $50$ runs of the same experiment, where the gateway node forwards the data between Node A and Node B. However, we also performed some continuous tests where the gateway node exchanges data back and forth between the two groups and found consistent results (i.e., the total time and energy consumption of the continuous tests were simply the sum of the time and energy for the isolated tests).  

%Additionally, we want to compare how different techniques can impact these times, and more specifically whether there are any gains, in either time or energy, to be had. 
%The techniques we are focusing on are ones where the gateway node is able to remain connected to both groups. 
%In the interest of fairness we have limited the comparisons to the {\it LCGM} scenario. %and have included our results in figures~\ref{fig:EnLCGM} and~\ref{fig:TimeLCGM}. 
%These techniques include the UDP method mentioned above and a method which we refer to as {\it Cut and Run}, where the gateway maintains both a WiFi connection and a WiFi Direct connection and then disconnects the WiFi connection to forward the data onto the second
%group.
%These techniques are then compared to a modified version of Android. This modified version does not have the static IP address range that the stock version of Android has, rather it changes the third byte to be different from the $49$ that is traditionally included in the operating system.

\vspace{-1.5mm}
\subsection{Numerical Results - Time Sharing}
\vspace{-1mm}
\begin{figure}
    \centering \includegraphics[width=\figwww]{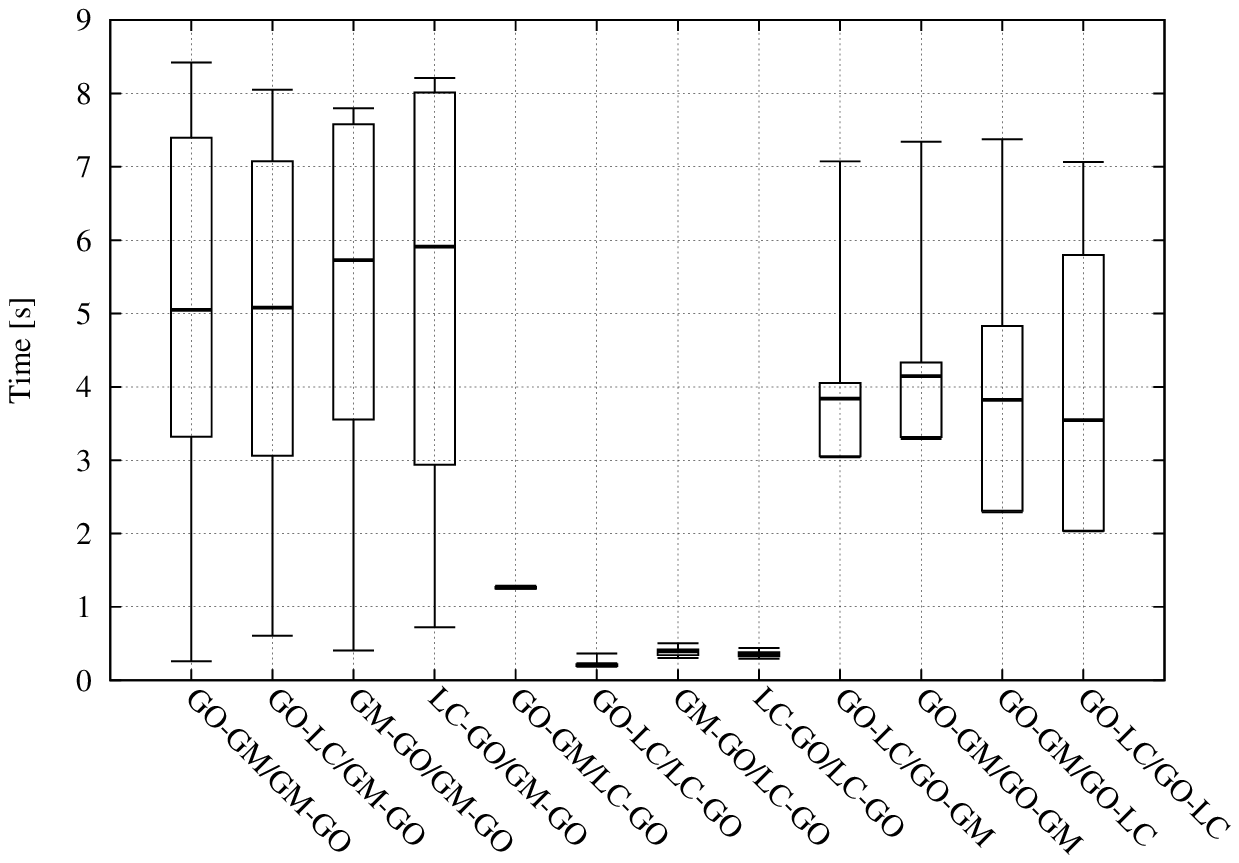}
    \caption{Experimental Measurements. Time required to switch between two groups for the different scenarios described in Section~\ref{WiFiDirectMulti}.}
    \label{fig:time}
\end{figure}

\begin{figure}
    \centering \includegraphics[width=\figwww]{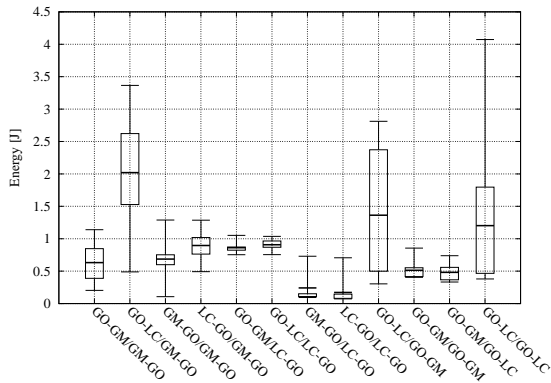}
    \caption{Experimental Measurements. Energy required to switch between two groups for the different scenarios described in Section~\ref{WiFiDirectMulti}.}
    \label{fig:energy}
\end{figure}

To measure the amount of energy and time required to switch between the two groups, we performed isolated tests to determine the impact that the switching process has on a particular node. We first established some baseline measurements for the transmission and reception energy consumptions. Using the experimental setup described above and Iperf~\cite{iperf}, we measured the average energy required to send (and receive) $10$~MB of data between devices, when acting as a GO, a GM or a LC. 
We then logged the total energy required at the gateway node to receive the $10$~MB data, switch and transmit the data to the other group. Finally, we subtracted the reception and transmission energy from the total energy to determine the energy required for switching between the groups. %An example of the power readings for a single experiment is provided in Figure~\ref{fig:sample}.

In Figures~\ref{fig:time} and~\ref{fig:energy}, we plot the time and energy, respectively, required to switch between the two groups, for all the scenarios described in Section~\ref{WiFiDirectMulti}. Figure~\ref{fig:time} shows that the switching time depends only on the method used by the gateway node for connecting to the second group. In particular, the lower switching times are achieved by the gateway that connects to an existing group as a LC. This is because the Android API that manages the standard WiFi connection allows to scan and connect to an existing AP faster than the WiFi Direct API, that instead relies on a combination of asynchronous call and event notifications. As a result, when connecting as a GM, the gateway is either able to join the second group in less than $500$ms (see the minimum values of cases */GM-GO in Figure~\ref{fig:time}) or, in other cases, the switching process can require even more than $8$s. We note that these variable connection times are due to the WiFi Direct protocol that requires, during the group formation process, an initial discovery phase where each device iteratively searches and listens over channels $1$, $6$ and $11$ for nearby clients~\cite{WiFip2p}~\cite{Mur2013}. 

When the gateway creates the second group and acts as a GO, instead, the average switching time is around $4$s, with lower connection times for the case in which it is connecting to a LC. It is important to notice that when the gateway node acts as a GO for the second group, it cannot send connection requests to the LC. In this case, instead, the LC is constantly scanning for the second group and, as soon as the gateway creates the group, it will connect to the group.

Our results are in line with the experimental evaluations of the WiFi Direct protocol presented in~\cite{Mur2013} and~\cite{Conti2013}. For example,~\cite{Conti2013} showed that the average time required for a device to autonomously create a group and immediately become a GO (autonomous group formation) is $3$s, while the average time required for a node to join an existing group is $6$s.

Regarding the energy, Figure~\ref{fig:energy} shows that the energy required to switch between two groups is similar across the different scenarios, with the exception of the cases in which the gateway is required to switch from a LC to a WiFi Direct node, e.g., a GO or a GM. This is due to the fact that the Android WiFi Direct implementation is built on top of the standard WiFi APIs but it requires some additional initialization (i.e, starting the WiFi Direct services) before the device can actually use the WiFi Direct protocol. For the same reason, switching between WiFi Direct roles or from a WiFi Direct role to a LC, instead, does not provide any significant impact on the energy consumption of the node. In addition, we note that even if the energy consumed by the gateway node when switching as a LC is comparable and, in some cases, lower than the energy required by the other scenarios, this energy is consumed in a short amount of time (i.e., around $500$ms in Figure~\ref{fig:time}) and thus the power consumption of this operation is much higher. This is because the LC implements a more aggressive approach for scanning and connecting to an existing group, that causes the higher power consumption when compared to the other methods.

\vspace{-1mm}
\subsection{Numerical Results - Simultaneous Connections}
\begin{figure}
    \centering \includegraphics[width=0.85\columnwidth]{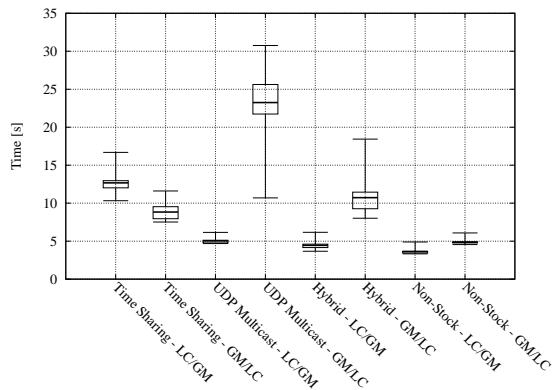}
    \caption{Experimental Measurements. Time required to transfer $10$ MB of data between two groups for a gateway node acting as LC and GM.}
    \label{fig:TimeLCGM}
\end{figure}

\begin{figure}
    \centering \includegraphics[width=0.85\columnwidth]{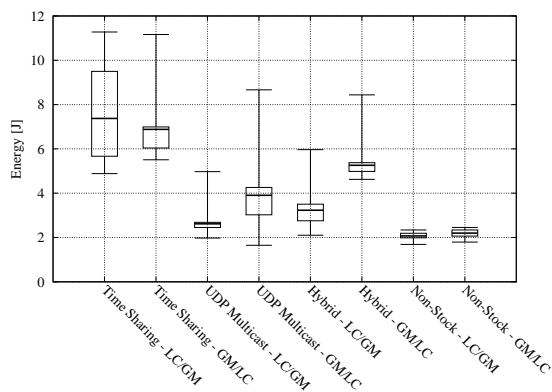}
    \caption{Experimental Measurements. Energy required to transfer $10$ MB of data between two groups for a gateway node acting as LC and GM.}
    \label{fig:EnergyLCGM}
\end{figure}

\vspace{-1mm}
We now focus on the gateway node configurations that allow for simultaneous connections in two WiFi Direct groups. In this regard, in Figures~\ref{fig:TimeLCGM} and~\ref{fig:EnergyLCGM} we plot the time and energy, respectively, required to transfer $10$ MB of data between the two groups, for the scenarios in which the gateway node acts as a LC in one group and a GM in the other  (i.e., scenarios GO-LC/GM-GO and GO-GM/LC-GO). We remind the reader that in these configurations (except for the {\it Time Sharing} scenarios that are included for comparison), the gateway node is allowed to maintain a simultaneous physical connection to both groups (see Section~\ref{sec:proposed_solutions}). Figures~\ref{fig:TimeLCGM} and~\ref{fig:EnergyLCGM} show that substantial gains in both time and energy can be achieved by the techniques that allow for simultaneous connection in both groups. As expected, the best performances are attained by the non-stock Android implementations, which represent the lower bound on the performance achievable by any scheme. Nevertheless, both the {\it UDP Multicast} and the {\it Hybrid} approaches perform very close to the lower bound, with the exception of the {\it GM/LC} configuration that requires a higher time and energy consumption. This is due to the encapsulation of a one-to-many unicast communication protocol that impacts the data reception over the WiFi Direct link of the {\it UDP Multicast} approach, and to the configuration switch of the {\it Hybrid} scheme. %requires some additional coordination and overhead. 
Moreover, we note that while the {\it Time Sharing}, {\it Hybrid} and {\it Non-Stock} implementations use TCP sockets for reliable data transmission, the {\it UDP Multicast} communication does not implement any retransmission mechanism %to assure reliability 
and, as such, is subject to a variable data loss (in our experiments, an average of $93\%$ of the total data was successfully delivered). 

Finally, while all the results presented in this section were obtained using a 2013 Nexus 7~\cite{Nexus7}, similar conclusions can be drawn when changing to a different device. In particular, we ran the tests presented in this section on a 2012 Nexus 7, which has a completely different hardware configuration (e.g, different memory, processor and wireless chipset), with Android $4.4.2$ and found similar relative results in term of transmission/reception and switching time, but an  overall higher (almost doubled) energy consumption for all schemes. These results suggest that similar conclusions can be drawn for different Android devices.

\section{Conclusions}
\label{sec:conclusions}

In this paper, we propose and explore different methods for allowing the communication among devices that belong to different WiFi Direct groups, in order to create multi-hop ad hoc networks. 
We present the limitations of enabling WiFi Direct multi-group communication, and we propose and analyze different mechanisms to allow the communications between devices belonging to different groups. For all the proposed methods, we discuss the achievable tradeoffs in terms of both time and energy.

Our experimental results show that a faster switching time can be achieved by connecting to an existing group as a legacy client, at the expense of a higher power consumption. In addition, switching from standard WiFi to WiFi Direct entails a higher energy consumption, when compared to all the other scenarios. Better performance can be achieved by techniques that exploit the device ability to maintain simultaneous physical connections to two groups. In particular, the {\it Hybrid} approach, where a UDP multicast communication is used as a control channel for triggering a configuration switch, allows for a reliable data transfer between groups and attains performance similar to a non-stock Android implementation.

%Our experimental results show that we can reduce the energy and time required to transfer data between groups by using a variety of techniques. Although these gains can be found with using the stock operating system, modifying the operaing system gave the best overall performance since the two stock methods either 1) were not able to transmit data in a bi-directional fashion or 2) exhibit drastically different performance in one direction when compared to the other. Additionally, by modifying the operating system we are able to continue to benefit from the reliability of TCP. 

Future work includes the inclusion of a routing protocol, and further modifications to Android OS to allow simultaneous communication over multiple WiFi Direct interfaces.

%Our experimental results show that a faster switching time can be achieved by connecting to an existing group as a legacy client, at the expense of a higher power consumption. In addition, switching from standard WiFi to WiFi Direct entails a higher energy consumption, when compared to all the other scenarios. Finally, our results suggest that when switching multiple times back and forth between two groups, the best performance is achieved by a gateway node that acts either as a legacy client or as a group member in both groups.

%Future work includes a comparison of these proposed techniques with a UDP flooding based protocol similar to the one proposed in~\cite{Duan2014}, as well as a technique that modifies the Android operating system to create multi-group WiFi Direct networks.
%A comparison of the proposed techniques with the performance achievable by reprogramming the Android operating system is considered as future work.

\vspace{-1mm}

\vspace{-0.6mm}
\bibliographystyle{IEEEtran}
\bibliography{IEEEabrv,main}

\end{document}